\documentclass{article}

\usepackage{arxiv}

\usepackage[utf8]{inputenc} 
\usepackage[T1]{fontenc}    
\usepackage{hyperref}       
\usepackage{url}            
\usepackage{booktabs}       
\usepackage{amsfonts}       
\usepackage{nicefrac}       
\usepackage{microtype}      
\usepackage{lipsum}
\usepackage{float}
\usepackage{mathtools}
\usepackage{bm}
\usepackage{amssymb}
\usepackage{longtable}

\title{Cellular State Transformations using Generative Adversarial Networks}

\author{
  Colin Targonski \\
  Department of Electrical and Computer Engineering\\
  Clemson University\\
  Clemson, SC 29634 \\
  \texttt{ctargon@clemson.edu} \\
   \And
 Benjamin T. Shealy \\
  Department of Electrical and Computer Engineering\\
  Clemson University\\
  Clemson, SC 29634 \\
  \texttt{btsheal@clemson.edu} \\
  \And
 Melissa C. Smith \\
  Department of Electrical and Computer Engineering\\
  Clemson University\\
  Clemson, SC 29634 \\
  \texttt{smithmc@clemson.edu} \\
  \And
  F. Alex Feltus\thanks{Corresponding author.}\\
  Department of Genetics \& Biochemistry\\
  Clemson University\\
  Clemson, SC 29634 \\
  \texttt{ffeltus@clemson.edu} \\
}

\begin{document}
\maketitle

\begin{abstract}
We introduce a novel method to unite deep learning with biology by which generative adversarial networks (GANs) generate transcriptome perturbations and reveal condition-defining gene expression patterns.  We find that a generator conditioned to perturb any input gene expression profile simulates a realistic transition between source and target RNA expression states.  The perturbed samples follow a similar distribution to original samples from the dataset, also suggesting these are biologically meaningful perturbations. Finally, we show that it is possible to identify the genes most positively and negatively perturbed by the generator and that the enriched biological function of the perturbed genes are realistic.  We call the framework the Transcriptome State Perturbation Generator (TSPG), which is open source software available at \url{https://github.com/ctargon/TSPG}. 
\end{abstract}

\keywords{Generative Adversarial Networks \and Bioinformatics \and Adversarial Examples}

\section{Introduction}
RNA sequencing (RNAseq) uses the capabilities of high-throughput DNA sequencing methods to provide insight into the transcriptional state of a biological sample \cite{kukurba2015rna}. An RNAseq gene expression profile can be coupled with metadata describing the sample, thus placing it in a biological context such as tissue source, aberrant phenotype such as disease, and other variables such as age, sex, medical history, and environmental factors. Traditional methods of high dimensional analysis are challenging to use on RNAseq datasets because of the high feature to sample ratio and often high noise levels in biological systems.   Modern machine learning and deep learning approaches have begun to surmount these obstacles and are increasingly  used to cluster and classify gene expression profiles into meaningful groups that correspond with metadata labels \cite{RocheTCGA, chen2016gene, lin2017using}.  

Deep learning \cite{lecun2015deep} has had tremendous success in image \cite{krizhevsky2012imagenet, he2016deep} and natural language \cite{hinton2012deep, hermann2015teaching} processing tasks due to its ability to abstract high level features from highly dimensional and noisy datasets. Deep learning has also been incorporated as the underlying framework behind powerful generative models such as variational autoencoders (VAEs) \cite{kingma2013auto} and generative adversarial networks (GANs) \cite{goodfellow2014generative}.  GANs, a focus of this work, consist of a generator that captures a data distribution and a discriminator that estimates the probability that a sample came from the training data rather than the generator. As training progresses, the generator produces increasingly realistic data while the discriminator becomes more adept at distinguishing real from fake. GANs have surged in popularity in the computer vision field due to several works creating exceptionally realistic images \cite{isola2017image, karras2017progressive, brock2018large}. 

While deep learning has proven itself useful in a variety of fields, surprising vulnerabilities exist within trained models. Adversarial examples are inputs to neural networks that an attacker has designed with the intention to cause the model to make a mistake \cite{szegedy2013intriguing, goodfellow2014explaining}. The general anatomy of an adversarial example consists of a calculated perturbation that is added to an input, which is subsequently passed to a trained model. Adversarial examples are generated to "trick" a neural network into confidently choosing a target class. For example, image \(x\), classified as a horse by a neural network, can be subtly perturbed and classified by the same neural network as a human.  Previous works \cite{goodfellow2014explaining, carlini2017towards, liu2016delving} have led to Xiao \textit{et al} \cite{xiao2018generating} proposing AdvGAN, a framework to produce perceptually realistic images in an efficient manner using GANs. In this report, we adapt the AdvGAN framework to model the transition between tissue transcriptome states.

Within the field of molecular biology, a limited amount of work involving generative models has been proposed. Ghahramani \textit{et al} \cite{ghahramani2018generative} proposed a Wasserstein GAN \cite{arjovsky2017wasserstein} to integrate epidermal datasets by generating samples that cover the full diversity of cell types. Additionally, the authors use the generative model for both dimensionality reduction and to observe the effect of cell state perturbations on gene expression.  Dizaji \textit{et al} \cite{ghasedi2018semi} introduced a semi-supervised approach to generate gene expression profiles of target genes using landmark genes and GANs.  This work adapts the AdvGAN framework to analyze gene expression patterns during cellular state transformations by training a generator on a specific set of genes with a conditional target. 

\section{Methods}

\subsection{Datasets and Dataset Preparation}
This study uses two gene expression matrices (GEMs) containing RNAseq expression levels for human tissue and tumor samples. The first dataset tested is the Genotype-Tissue Expression (GTEx) dataset, which contains 11,688 samples across 56,202 genes representing 53 uniquely labeled tissue types. The second dataset considered is the Cancer Genome Atlas (TCGA) dataset \cite{pantcgacitation}, which contains 11,092 patients across 60,483 genes representing 33 unique tumor types of distinct disease progression. We normalize expression levels by scaling each feature (gene) between the range \([0,1]\).  We use the Broad Institute Molecular Signature Database gene subsets (MSigDB v6.2) (MSigDB) \cite{subramanian2005gene} to extract sub-GEMs from GTEx and TCGA of lower dimensionality in order to reduce genetic complexity and obtain more interpretable results.  We normalize the GEM data using the Python sci-kit learn package \cite{scikit-learn}. All neural network development and training is performed using Tensorflow \cite{tensorflow2015-whitepaper}. Training of the networks utilized the Clemson University Palmetto Cluster using Nvidia V100 GPUs.  The entire implementation is available on GitHub at \url{https://github.com/ctargon/TSPG}. 

\subsection{Transcriptome State Perturbation Generator (TSPG)}
The general framework of the GAN used in this work was adapted from the AdvGAN model introduced by \cite{xiao2018generating}, which was influenced and derived from previous models and criterion metrics from both the GAN research community and the adversarial attack community. To formulate the adversarial generation that TSPG aims to perform, let us consider an input pair \((x_{i}, y_{i})\) to be the \(i\)th input pair of a dataset consisting of feature vector \(x_{i} \in X\) generated according to an unknown distribution \(x_{i} \sim P_{data}\), where \(X \subseteq \mathbb{R}^{n}\) with \(n\) being the number of genes (features), and \(y_{i} \in Y\) the corresponding label. The goal of a classification system, which we will refer to as the target model, is to learn a mapping \(f : X \to Y \) from input domain to label domain. The goal of the adversary (in this case, the generator) is to learn to produce \(x_{adv}\) that is classified as \(f(x_{adv}) = t\) where \(t\) is a target (the targeted case) and different from the ground truth label \(y\).

\begin{figure}
\begin{center}
\includegraphics[scale=0.4]{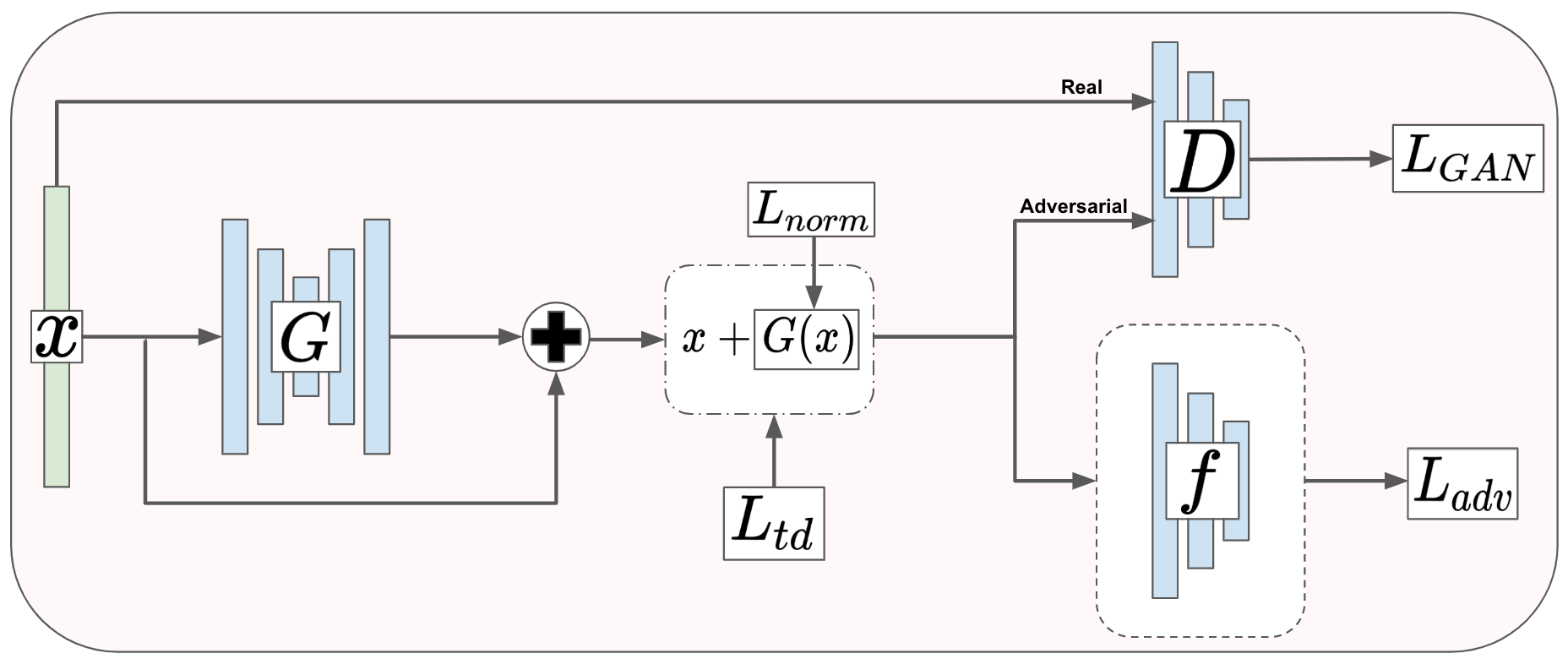}
\end{center}
\caption{Architecture of transcriptome state perturbation generator (TSPG).}
\label{fig:TSPGarch}
\end{figure}

The TSPG framework can be seen in Figure \ref{fig:TSPGarch} and contains a generator \(G\), a discriminator \(D\), and a pretrained target network \(f\). \(G\) takes an input vector \(x\), with the number of genes being the number of dimensions, and generates a perturbation \(G(x)\). \(G(x)\) is added to the original instance to create an adversarial example \(x_{adv} = x + G(x)\) which is then passed to networks \(D\) and \(f\). The goal of \(D\) is to encourage the generated example \(x_{adv}\) to be indistinguishable from data sampled from \(P_{data}\), which is quantified by \(L_{GAN}\) and shown below:
\begin{equation}
\label{LGAN}
    L_{GAN} = 
    \mathbb{E}_{x}[\text{log}D(x)]
    -\mathbb{E}_{x}[\text{log}(1 - D(x + G(x)))]
\end{equation}
For \(L_{GAN}\), we use the least squares objective proposed in \cite{mao2017least}, as it has been shown to stabilize training and boost results. The network \(f\) outputs a class prediction for input \(x_{adv}\) based on the pretrained weights. The goal of the generator is to produce adversarial example \(x_{adv}\) that is classified as \(t\) by \(f\). The loss for tricking \(f\) in such a manner, proposed by \cite{carlini2017towards}, is:
\begin{equation}
\label{Ladv}
    L^{f}_{adv} = \text{max}(\text{max}_{i \neq t}f(x_{adv})_{i} - f(x_{adv})_t, \kappa)
\end{equation}
where \(t\) is the target class, and we set confidence \(\kappa = 0\). This loss term encourages the generator to produce data that fools \(f\) into predicting \(t\). 

We also use two terms, \(L_{norm}\) and \(L_{td}\), which are used to stabilize training and encourage \(x_{adv}\) to fall within the target data distribution \(P_{t}\). \(L_{norm}\) is simply the \(L_{2}\) norm placed on the generated perturbation \(G(x)\):
\begin{equation}
\label{Lnorm}
    L_{norm} = \|G(x)\|_{2}
\end{equation}
This loss encourages the overall perturbation to be smaller, and we found this stabilizes training and produces better results. The final criterion for the loss term, \(L_{td}\), is a measure of \(L_{1}\) distance between an adversary example \(x_{adv}\) and a randomly sampled target vector \(r_{t}\), where \(r_{t} \sim \mathcal{N}(\mu_{t}, \Sigma_{t})\). \(\mathcal{N}(\mu_{t}, \Sigma_{t})\) is a normal distribution modeled with the mean genetic expression level \(\mu\) of target class \(t\) and covariance matrix \(\Sigma\) of target class \(t\). The loss is formulated as:
\begin{equation}
\label{Ltd}
    L_{td} = \vert x_{adv} - r_{t} \vert
\end{equation}
\(L_{td}\) encourages the generator to create data more tightly within the bounds of the target class distribution. Similar principles are used in \cite{isola2017image, zhu2017unpaired} to reduce the space of possible mappings and create sharper data instances. Together, the loss for \(G\) can be summed as:
\begin{equation}
\label{Ltot}
    L = L_{GAN} + L_{adv} + L_{norm} + L_{td}
\end{equation}

The architectures used by \(G\) and \(D\) are similar to those used in \cite{xiao2018generating}, which are influenced from image to image translation architectures \cite{isola2017image, zhu2017unpaired}. \(G\) adopts an hourglass shaped structure which contains three downsampling fully connected layers of sizes 512, 256, and 128, followed by three residual blocks, finishing with two upsampling layers of sizes 256 and 512 and then a layer of the same shape as the input dimension. The rectified linear unit (ReLU) activation function is used on each layer barring the final output, which uses the hyperbolic tangent (tanh). Each hidden dense layer is coupled with a batch normalization layer using default Tensorflow parameters. The tanh function allows the generator to produce a perturbation ranging from \([-1,1]\). After the \(x_{adv} = x + G(x)\) operation, \(x_{adv}\) is clipped to values between \([0,1]\), which are valid gene expression values. 

The discriminator uses a small architecture containing three dense layers of units 512, 256, and 128. Each dense layer is followed by a batch normalization layer and a leaky ReLU activation function with alpha set to \(0.2\). The target model used in the experiments contains three dense layers of size 1024, 512, and 128, each layer using the ReLU activation function. All three networks are trained using the Adam optimizer \cite{kingma2014adam} with learning rates 0.0002, 0.0001, and 0.001 for the generator, discriminator, and target model, respectively. The generator uses bootstrapping by randomly selecting 80\% of each class for training and holding out the remaining 20\% for testing.

\subsection{Data Visualization}
Once a generator is trained using the framework and a given target, an input vector was fed to the network for inference, resulting in a perturbation and newly spawned adversarial example. We chose to perturb 100 randomly chosen samples from a given GEM dataset, and then we plotted the results using t-SNE \cite{maaten2008visualizing} along with the samples from 10 randomly selected classes, including the target class. This created a low dimensional representation of the sample distributions, allowing us to observe what distribution the adversarial examples follow. We also provide methods to view the RNAseq changes in the form of a heatmap. The components involved in a targeted perturbation generation include a sample, a perturbation, and an adversarial example. Additionally, we used the mean vector of a target class to visually inspect how similar an adversarial example is to the target class. With these ingredients, we can produce a plot depicting a gene-wise transformation of an input cell to a target cell, including the RNAseq expression level differences.

\subsection{Gene Set Functional Profiling}
The top and bottom twenty most highly perturbed genes were input into the ToppFun tool (\url{https://toppgene.cchmc.org}) \cite{chen2009toppgene} in March 2019.  Occasionally, the HGNC symbol was not recognized for 1/20 genes in the list. Functional labels were considered significant at a q-value Benjamini and Hochberg corrected False Discovery Rate of q < 0.0001.

\section{Results}
\label{sec:results}
We first examined the raw adversarial generation capabilities using 50 "Hallmark" gene sets defined in MSigDB. Table \ref{table:adversarialaccuracy}  contains the results for various Hallmark gene sets being used as feature input to the TSPG framework, as well as the corresponding target class. The furthest right column, \textit{f Accuracy}, represents the target model accuracy when the input is perturbed \textit{from any input source} (i.e. all samples from the dataset) and the label is the target class. Thus, the accuracy represents the ability of the generator to "trick" the target model into classifying a sample as a specified class.


\begin{table}[H]
\caption{Accuracy of target network on perturbed datasets.}
\label{table:adversarialaccuracy}
\centering
 \begin{tabular}{l  c  l  c} 
 \hline
 Gene Set & Num Genes & Target Class & $f$ Accuracy \\ [0.5ex] 
 \hline\hline
 Hallmark Hedgehog Signaling & 36 & Nerve - Tibial & 100\%  \\ 
 Hallmark Peroxisome & 107 & Brain - Spinal Chord & 100\% \\
 Hallmark Apoptosis & 161 & Lung & 99.9\% \\
 Hallmark E2F Targets & 200 & Artery - Coronary & 99.8\% \\
 Hallmark All & 4386 & Thyroid & 100\% \\
 Hallmark All & 4386 & Heart - Left Ventricle & 100\% \\
 \hline\hline
\end{tabular}
\end{table}

\begin{figure}
\begin{center}
\includegraphics[width=\linewidth, clip]{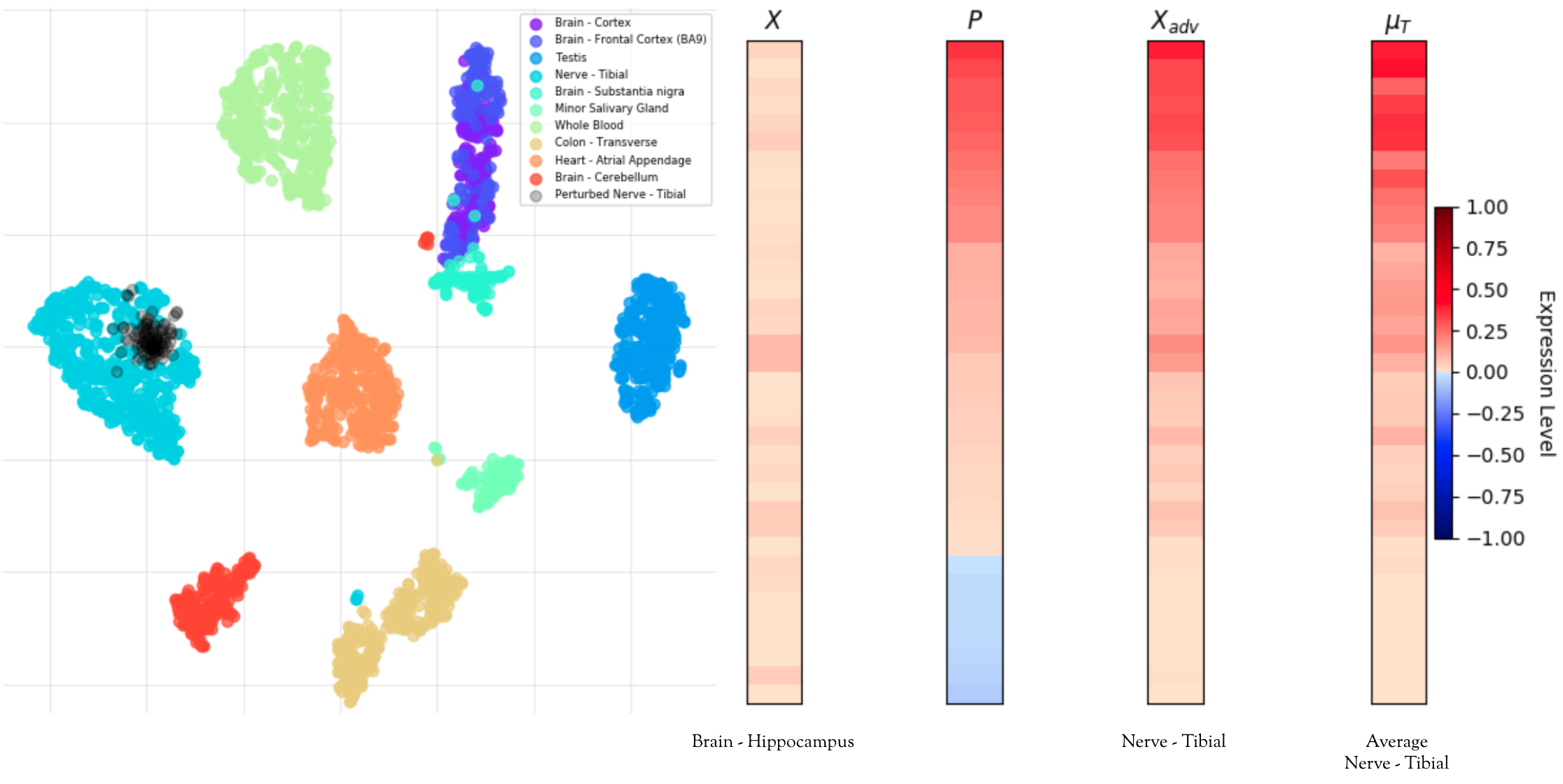}
\end{center}
\caption{\textbf{Adversarial generation for Nerve-Tibial target using the Hallmark Hedgehog Signaling gene set.}  t-SNE plot of original and perturbed samples using the Hallmark Hedgehog Signaling gene set (left). Heatmap of cellular transformation from Brain-Hippocampus to Nerve-Tibial (right). Perturbation ($P$) ranges from $[-1,1]$, which is added to original sample ($X$), then adversarial example ($x_{adv}$) is clipped to $[0,1]$.  The mean expression vector ($\mu_{T}$) of the target class (Nerve-Tibial) is shown.}
\label{fig:hhTSPG}
\end{figure}

The high classification accuracy suggests that the generator is powerful enough to exploit the decision boundary the target model defined no matter what class the input came from. While this is in itself interesting, we still did not know what distribution the adversarial samples might follow. This left us with three questions: 1) Will generated adversarial examples follow the distribution of the target class? 2) Are the perturbations created by the generator nonsensically changing expression values to satisfy the criterion of tricking a target model? 3) Do the most highly perturbed genes encode biological function of the target condition?

To test if the adversarial examples exhibited similar high dimensional structure as the original target samples, we used the dimensionality reduction and visualization tool t-SNE \cite{maaten2008visualizing}. This technique allowed us to ask if perturbed samples clustered with the target condition distribution. In Figure \ref{fig:hhTSPG}, ten different classes from the GTEx dataset were plotted, as well as 100 randomly chosen adversarial examples that were created from varying source tissues in the original dataset. Each sample was originally 36 dimensions, representing the 36 genes in the Hallmark Hedgehog Signaling gene set. We observed that the original (unperturbed) Nerve-Tibial samples (dark cyan) and the adversarial (perturbed) Nerve-Tibial samples (black) formed a cluster on the left side of the plot. The adversarial (black) points were originally 100 different samples from the GTEx dataset of varying class description, then were passed through the generative process as described in Figure \ref{fig:TSPGarch}, thus transforming these points to the target condition. It is evident from Figure \ref{fig:hhTSPG} that the perturbed samples follow a similar data distribution as the original target condition.

The right hand side of Figure \ref{fig:hhTSPG} shows the step by step process of generating an adversarial example. In this case, $x$ was an original sample from the GTEx dataset of class type Brain-Hippocampus. $x$ was passed through the generator to produce $P$, which was then added to $x$ to create adversarial example $x_{adv}$. $P$ ranges between $[-1,1]$ to allow the generator to entirely "activate" or "silence" any gene. $x_{adv}$ was clipped to valid expression values in the range of $[0,1]$. This generated example $x_{adv}$ was then classified as target class $t$ by a pretrained model (the target model). What was left are the perturbation values $P$ describing the exact gene expression changes, on a per gene basis, that occurred in order to transform, in this instance, Brain-Hippocampus tissue to Nerve-Tibial tissue. Additionally, we showed $\mu_{t}$, the mean expression vector for the target class (Nerve-Tibial in this example) for direct visual comparison between an adversarial example $x_{adv}$ and the target class.

\begin{figure}
\centering
\includegraphics[width=\linewidth]{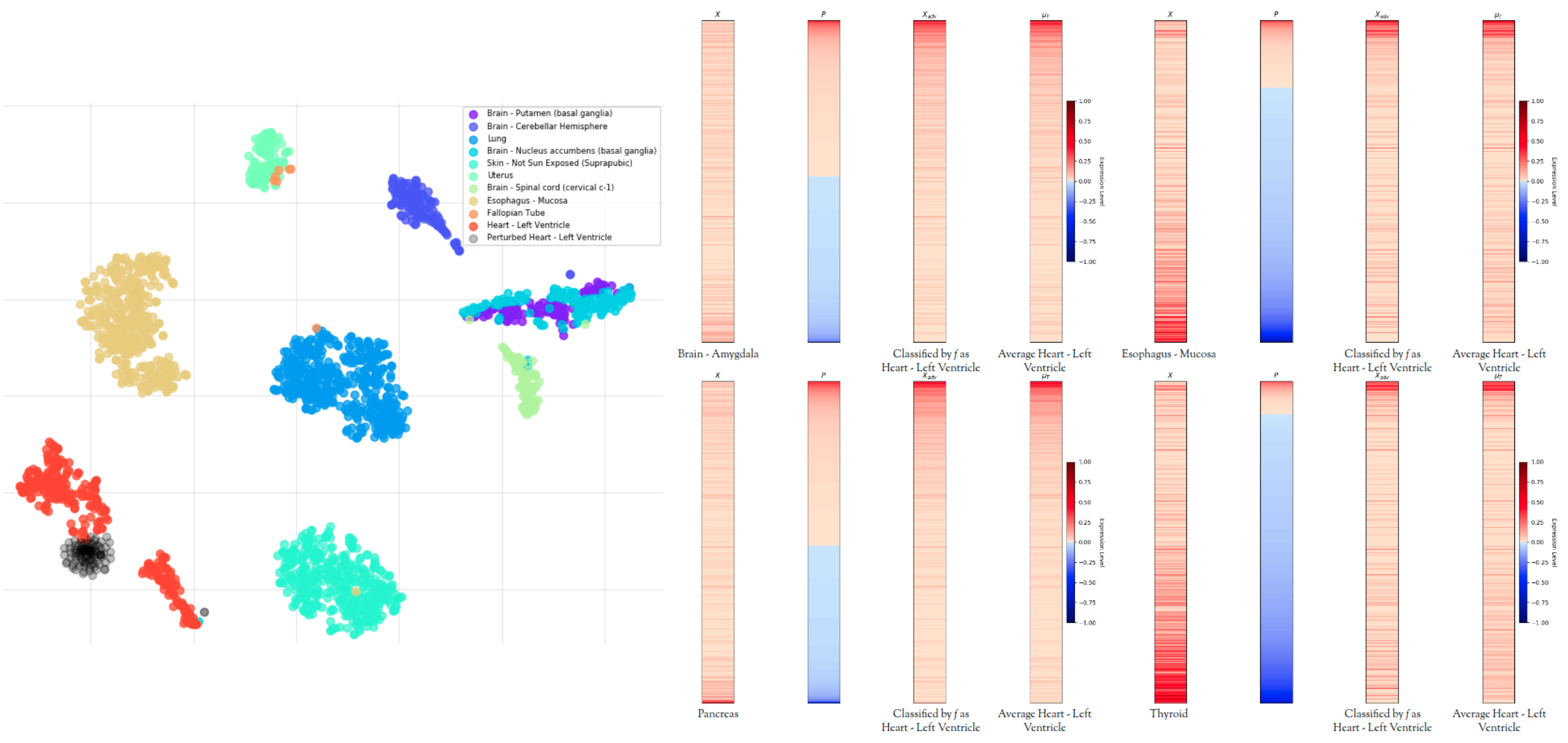}
\caption{\textbf{Adversarial generation for Heart-Left Ventricle target using all Hallmark genes as the input gene set.}  t-SNE plot of original and perturbed samples using the all Hallmark genes (left). Heatmap of cellular transformations from Brain-Amygdala, Esophagus-Mucosa, Pancreas, and Thyroid to to Heart-Left Ventricle (right). Perturbations ($P$) range from $[-1,1]$, which is added to original sample ($x$), then adversarial example ($x_{adv}$) is clipped to $[0,1]$.  The mean expression vector ($\mu_{T}$) of the target class (Heart-Left Ventricle) is shown.}
\label{fig:allheartTSPG}
\end{figure}

\begin{figure}
\centering
\includegraphics[width=\linewidth]{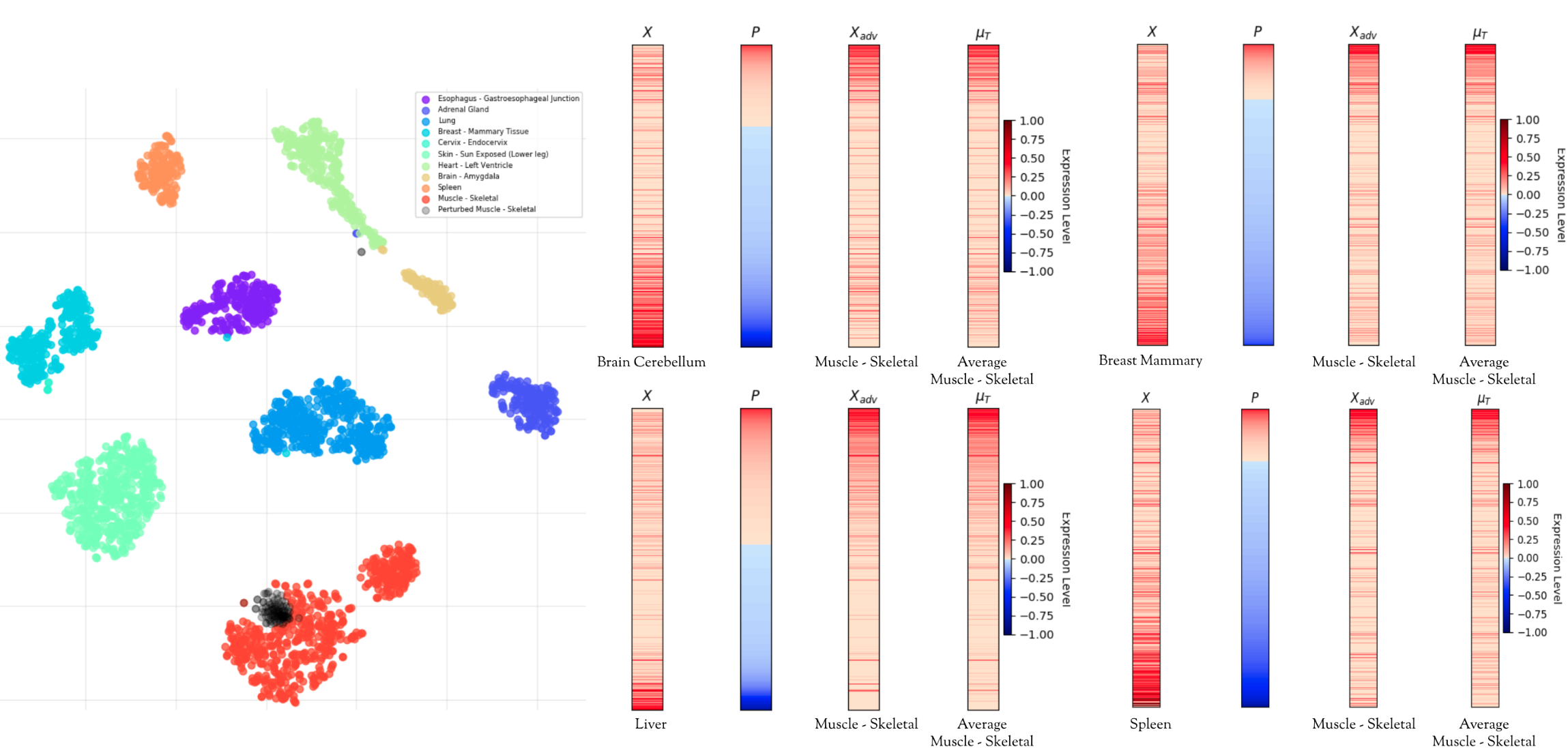}
\caption{\textbf{Adversarial generation for Muscle-Skeletal target using all Hallmark genes as the input gene set.}  t-SNE plot of original and perturbed samples using the all Hallmark genes (left). Heatmap of cellular transformations from Brain-Cerebellum, Breast-Mammary, Liver, and Spleen to to Muscle-Skeletal (right). Perturbations ($P$) range from $[-1,1]$, which is added to original sample ($x$), then adversarial example ($x_{adv}$) is clipped to $[0,1]$.  The mean expression vector ($\mu_{T}$) of the target class (Muscle-Skeletal) is shown.}
\label{fig:allTSPG}
\end{figure}

We also include Figures \ref{fig:allheartTSPG} and \ref{fig:allTSPG} as examples of TSPG working with a larger input gene set where all 4,386 Hallmark genes were combined. In this case, the perturbed samples clustered with original samples of both target classes (Heart - Left Ventricle and Muscle-Skeletal). Additionally, if we visually inspect the similarity between $x_{adv}$ and $\mu_{T}$ in each heatmap, the expression vectors appear to be similar. This experiment demonstrates the ability of TSPG to generalize to variably sized input gene sets. 

The classification accuracy, t-SNE visualizations, and perturbation heatmap results suggest that the perturbation of the hallmark gene expression patterns shift the state of the source tissue to the target tissue. To test if the perturbed genes are at all relevant to the source or target tissue, we randomly selected a sample from each of the 53 tissues available in the GTEx dataset and passed it through the generator that was trained with Heart, Left Ventricle being the target. Then, we took the 20 most \textit{positively perturbed} genes (genes being "turned on" or expressed) and the 20 most \textit{negatively perturbed} genes (genes being "turned off" or silenced) for each of the 53 state transitions. Interestingly, only 85 out of 4,386 genes were found in the top 20 most perturbed list for each of the 53 GTEx experiments suggesting that the same genes tend to be perturbed towards the target tissue.  On the other hand, the twenty most negatively perturbed genes summed to 642 unique genes suggesting that a greater variety of genes must be "turned off" depending on the phenotype of the source tissues. 

\begin{table}
\caption{Enriched Biological Functions of the Top Twenty Most Positively Perturbed Genes Leading to the Target Condition Heart-Left Ventricle}
\label{table:toptoppfun}
\centering
 \begin{tabular}{p{3cm}  p{1cm} p{9cm}  p{2cm}}
 \hline
Term Category & Enriched Terms* & First Hit Term Description &	q-value \\ [0.5ex]
 \hline\hline
Coexpression & 933 & Rat Breast\_Giusti09\_300genes & 1.72E-25 \\ 
Coexpression Atlas & 985 & Heart muscle & 8.01E-22  \\
Computational & 315 & Neighborhood of MYL2 & 5.29E-22  \\
Disease & 969 & Cardiomyopathy, Familial Idiopathic & 1.47E-09  \\
Drug & 1522 & Doxorubicin & 2.88E-11  \\
Gene Family & 22 & F-type ATPases|Mitochondrial complex V: ATP synthase & 2.28E-11  \\
GO: BP & 3311 & Striated muscle contraction & 2.01E-15  \\
GO: CC & 852 & Sarcomere & 2.53E-15  \\
GO: MF & 225 & Structural constituent of muscle & 8.08E-06  \\
Human Phenotype & 867 & Sudden death & 2.04E-07  \\
Interaction & 280 & VDAC1 interactions & 1.94E-05  \\
MicroRNA & 1 & Hsa-miR-610:mirSVR\_nonconserved\_lowEffect-0.1-0.5 & 3.77E-05  \\
Mouse Phenotype & 1419 & Cardiac hypertrophy & 2.47E-09  \\
Pathway & 1340 & Cardiac muscle contraction & 4.01E-14  \\
Pubmed & 10774 & Clinical features and outcome of hypertrophic cardiomyopathy associated with triple sarcomere protein gene mutations. & 1.13E-19  \\
ToppCell Atlas & 209 & MouseAtlas Mouse Multiple Adult muscle Subtype Heart\_cardiac muscle cell & 2.90E-18  \\
TFBS & 21 & V\$MEF2\_02 & 8.40E-06  \\
 \hline\hline
\end{tabular}
\end{table}

We next asked if the 20 most positively perturbed genes out of the pool of 4,386 genes were enriched (B\&H FDR q < 0.0001) for biological function relevant to the target tissue (i.e. Heart-Left Ventricle). A representative enriched term for each of the 16 term categories and the total number of enriched terms across all 53 experiments is shown in Table \ref{table:toptoppfun}. Clearly, many of the enriched terms are related to muscle and heart biology. Conversely, we asked if the 20 most negatively perturbed genes were related to the source tissue’s biology. Supplemental Table S1 shows the most significant PubMed scientific articles that were non-randomly associated with these genes. We found that the source tissue is related to the PubMed articles associated with these genes in most cases. All term enrichments can be explored in Supplemental Table S2.

\begin{figure}[H]
\centering
\includegraphics[width=\linewidth]{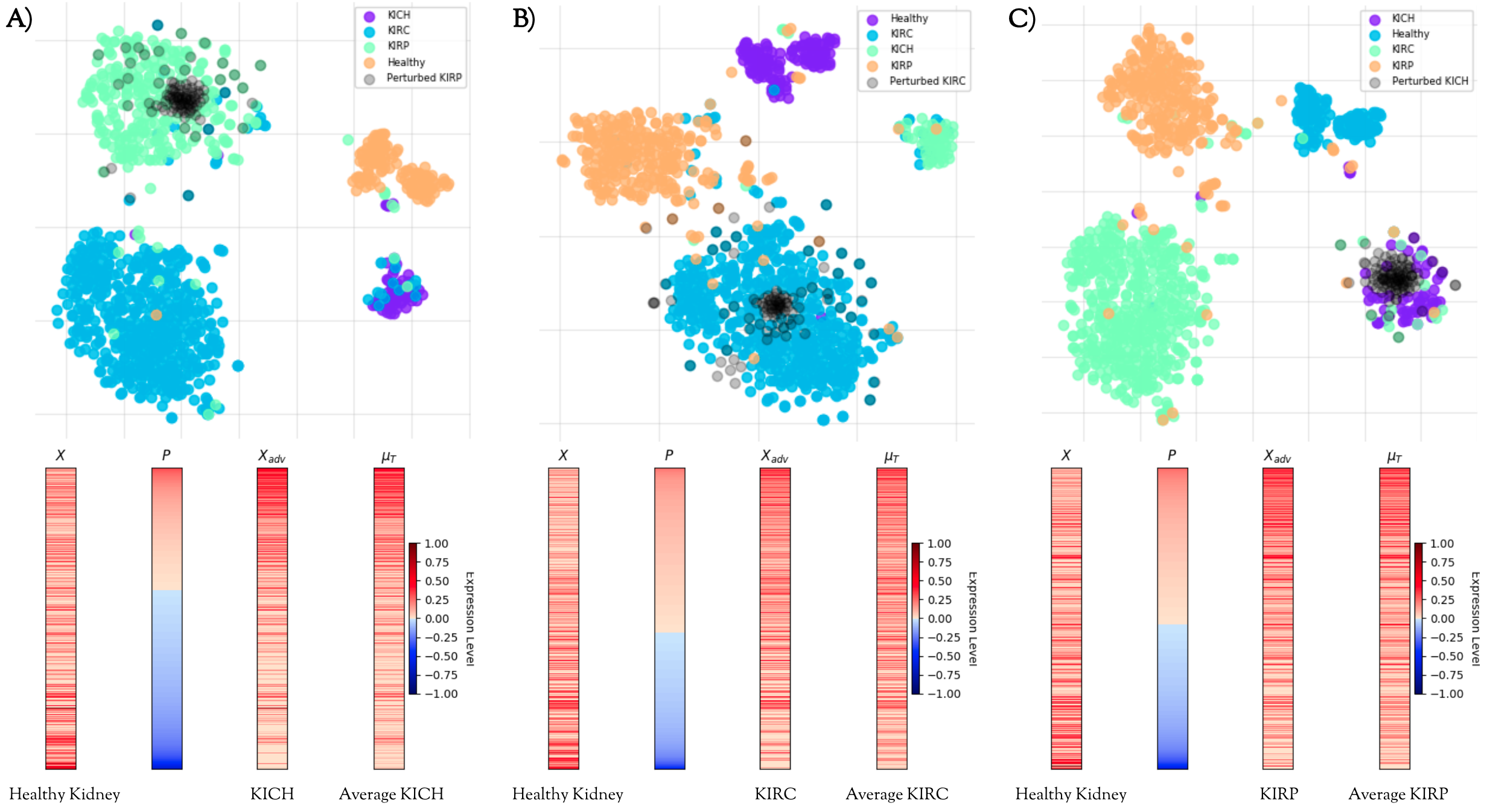}
\caption{\textbf{Adversarial generation for three subtypes of Kidney cancer using all Hallmark genes as the input gene set.} t-SNE plot and corresponding heatmap of cellular transformation from healthy to KICH (A), healthy to KIRC (B), and healthy to KIRP (C). Perturbations ($P$) range from $[-1,1]$, which is added to original sample ($X$), then adversarial example ($X_{adv}$) is clipped to $[0,1]$.  The mean expression vector ($\mu_{T}$) of the target class is shown.}
\label{fig:kidneyallTSPG}
\end{figure}

In order to determine the efficacy of TSPG on more subtle transcriptome state transitions, we tested TSPG on samples from the same tissue of origin, namely healthy and cancerous kidney tissue extracted from the TCGA project. Using all 4,386 Hallmark genes as input, we examined the transcriptome transitions between healthy (“solid tissue normal”) into one of three renal tumor subtypes: clear cell renal carcinoma (KIRC), renal papillary cell carcinoma (KIRP), and chromophobe renal cell carcinoma (KICH) as shown in Figure \ref{fig:kidneyallTSPG}. For each transition between healthy and the three tumor sub-types, we identified the genes that were among the top 20 genes most positively and negatively perturbed during transition (Supplemental Table S2).  For the most negatively perturbed genes in each case, the genes tended to be enriched for top expressed genes in kidney tissue as defined by the GenitoUrinary Development Molecular Anatomy Project (GUDMAP) database \cite{harding2011gudmap}.  In contrast to the negatively perturbed genes appearing to have a normal kidney gene expression pattern, the top positively perturbed genes showed different enriched functions for each tumor subtype.  A notable enriched function in KIRC genes was an association with genes involved in hypoxia, a hallmark of tumor progression where tumors avoid senescence by increasing blood supply (e.g. MSigDB hallmark hypoxia gene set was enriched).  A notable function enriched in positively perturbed KIRP genes is an enrichment in membrane-binding annexin activity (e.g. \cite{rescher2004annexins}), a dysregulated function in cancer \cite{lauritzen2015annexins} including renal carcinoma \cite{yang2015annexin,carlini2017towards}. Finally, positively perturbed genes from healthy to KICH exhibited an enrichment for genes upregulated by the estrogen receptor alpha  (e.g. \cite{stein2008estrogen}), and estrogen responsive genes have been implicated in kidney cancer \cite{liu2015expression}.  These results suggest that negatively perturbed genes  exhibit similar functions when transitioning to any RCC subtype, but the positively perturbed genes encode collective function unique to each subtype. Simply put, negatively perturbed genes are related to source tissue (i.e. healthy kidney) functionality, while positively perturbed genes relate to functionality specific to the target tissue (i.e. three cancerous kidney subtypes).

\section{Discussion}
Our preliminary experiments confirmed that neural networks used for classification of RNAseq data are indeed vulnerable to adversarial attacks. The results of Table \ref{table:adversarialaccuracy} show that it is actually quite trivial for the generator to "trick" the target model into a specific classification no matter what the input is. This is likely due to the fact that the target model is relatively simple and was not trained using any defensive strategies against attacks. It is noteworthy that \cite{szegedy2013intriguing} found that the fundamental nature of adversarial examples do not differ across models trained on different subsets of data or models that had varying hyperparameters (number of layers, number of neurons in each layer, weight normalization schemes, etc.). Thus, we decided it was not necessary to test across different MLPs. Additionally, a noteworthy standard for adversarial attacks in image processing is that the perturbation is small in magnitude (equation \ref{Lnorm}); ideally, the perturbation to an input is not noticeable to the human eye. While this criterion is foundational for adversarial attacks in the computer vision domain, we are more interested in making \textit{meaningful} transitions from source to target states, not in meaninglessly fooling a neural network. Therefore, despite including equation \ref{Lnorm} to stabilize training, it was not imperative that the perturbations were small in magnitude. A large shift in gene expression may be required to transition between two disparate tissue states.

Due to the remarkable success of GANs in various data intensive applications, the objective of fooling a MLP was expected to be accomplished trivially. Therefore, we decided to take the adversarial attack paradigm several steps forward and demonstrated that generative adversarial networks produce biologically meaningful perturbations when trained under the proper criteria. We adapted and added criteria to the AdvGAN framework, which originated from within the computer vision community, to produce meaningful state transitions. While we used equation \ref{Lnorm} to limit the magnitude of the perturbation, we found that adding equation \ref{Ltd} to the generator loss function significantly improved the realism of the results with respect to t-SNE plots and heatmap results.

What do these results mean for the study of biological systems?  While the results of this analysis pointed to the near perfect adversarial attack rates (Table 1) and clear correspondence of the perturbed gene expression patterns to the target condition (Figure \ref{fig:hhTSPG}-\ref{fig:kidneyallTSPG}), we were concerned that we had nonsensically transformed biological states. However, the strong signal of biological function (Table 2, Table S2) and continued selection of the same positively perturbed genes leading to the target gene expression pattern suggests that the algorithm shifts the expression patterns in a biologically meaningful way.  Thus, TSPG could be a novel method to detect genetic subsystems that are responsible for differentiating between both differing tissues as well as wild type and aberrant tissue states. Furthermore, generative models, particularly GANs, have been sparingly used in the field of biology. This can be attributed to a multitude of factors including the limited number of adequate datasets and the unnatural structure of RNAseq data  when compared to images and language waveforms. This work serves as a proof of concept that GANs can be applied to RNAseq data to uncover meaningful gene expression patterns between different human tissue samples.

\section{Acknowledgments}
This work was supported by National Science Foundation Award \#1659300 "CC*Data: National Cyberinfrastructure for Scientific Data Analysis at Scale (SciDAS)".  Clemson University is acknowledged for generous allotment of compute time on the Palmetto cluster. We would also like to thank Sufeng Niu and Courtney Shearer for the valuable discussions we had regarding this work.

\bibliographystyle{unsrt}  
\bibliography{references}  
\end{document}